\documentclass[preprint]{aastex}

\slugcomment{ }

\shorttitle{Blind HI Survey of the M81 Group}
\shortauthors{Boyce et al.}

\begin{document}


\title{A Blind HI Survey of the M81 Group}


\author{P. J. Boyce\altaffilmark{1}, 
 R. F. Minchin\altaffilmark{2}, V. A. Kilborn\altaffilmark{3}, 
 M. J. Disney\altaffilmark{2}, 
 R. H. Lang\altaffilmark{2}, 
 C. A. Jordan\altaffilmark{3}, 
 M. Grossi\altaffilmark{2}, 
 A. G. Lyne\altaffilmark{3}, 
 R. J. Cohen\altaffilmark{3}
  I. M. Morison\altaffilmark{3} \and S. Phillipps\altaffilmark{1}}


\altaffiltext{1}{Astronomy Group, Department of Physics, University of Bristol, Tyndall
 Avenue, Bristol, BS8 1TL; Peter.Boyce@bristol.ac.uk}
\altaffiltext{2}{University of Wales, Cardiff, Department of Physics and 
 Astronomy, P.O. Box 913, Cardiff CF2 3YB}
\altaffiltext{3}{University of Manchester, Jodrell Bank Observatory, Macclesfield, 
 Cheshire, SK11 9DL}


\begin{abstract} 
 Results are presented of the first blind HI survey of the M81 group of galaxies. 
 The data were taken as part of the HI Jodrell All Sky Survey (HIJASS). 
 The survey reveals several new aspects to the complex morphology of 
 the HI distribution in the group. All four of the known dwarf 
 irregular (dIrr)  
  galaxies close to M81 can be unambiguously seen in the HIJASS data. 
  Each forms part of the complex tidal structure in the area. We suggest that 
 at least three of these galaxies may have formed recently from the 
 tidal debris in which they are embedded. The 
 structure connecting M81 to NGC2976 is revealed as a single  
 tidal bridge of mass $\simeq$2.1$\times$10$^{8}$~M$_{\odot}$ 
  and 
 projected spatial extent $\simeq$80~kpc. Two `spurs' of HI projecting from 
 the M81 complex to lower declinations are 
 traced over a considerably larger spatial and velocity extent than 
 by previous surveys. The dwarf elliptical (dE) galaxies BK5N and Kar~64 lie at the spatial
  extremity of 
 one of these features and appear to be associated with it. 
  We suggest that these may be the remnants of  dIrrs 
   which have been stripped of gas and transmuted into dEs by close 
 gravitational encounters with NGC3077. The nucleated dE galaxy Kar~61 is 
 unambiguously detected
  in HI for the first time and has 
 an HI mass of $\sim$10$^{8}$~M$_{\odot}$, further confirming it as a dE/dIrr 
 transitional object.  HIJASS has revealed one new possible group member, 
 HIJASS~J1021+6842. This object contains 
 $\simeq$2$\times$10$^{7}$~M$_{\odot}$ of 
 HI and lies $\simeq$105\arcmin\  from 
 IC2574. It has no optical counterpart on the Digital Sky Survey. 
\end{abstract}




\section{Introduction}

The M81 group of galaxies is one of the nearest groups to our own.  
 Hence detailed study of the group  may impact on many key 
 areas of astronomy, such as structure formation, galaxy evolution, 
 star formation and dark matter. The present catalogue of 
 group members is the result of several optical 
  surveys  \citep[e.g.][]{kar85,kar98,kar01}. 
 The group contains one large spiral galaxy (M81), two peculiar 
 galaxies (M82 and NGC3077), two small spirals (NGC2976, IC2574) 
  and a large number of dwarfs.  
 The core galaxies of the group are  strongly interacting: there is 
 a large dynamically complex HI cloud embedding M81, M82, NGC3077 and NGC2976
 \citep{app81,yua94}.

Several pointed HI surveys have been undertaken to search for HI in  
 M81 group candidate dwarf galaxies \citep{huc98,vd98,huc00}.
  There are about 23 dwarf members 
 or candidate members (12 dIrrs, 9 dEs, 2 BCDs). HI has been 
  unambiguously detected from 
   9 of the dIrrs and from 1 BCD. 
   Measured velocities for M81 group objects lie in the 
 range --140~km\,s$^{-1}$ to +350~km\,s$^{-1}$. 
 Dwarf galaxies are difficult to unambiguously detect in HI within the 
  HI complex around M81. 
   Elsewhere, HI from group members may be
   masked by the strong Galactic HI emission.

 In this paper we present results from the first blind HI survey of the 
 M81 group. The data for this were taken as part of the 
 HI Jodrell All Sky Survey (HIJASS) which is described in Section~2. 
  Section 3 presents the observational 
 results for the M81 group of galaxies. Section~4 discusses the results 
 and presents our conclusions. A distance to the M81 Group of D=3.63~Mpc \citep{fre94} is 
 assumed 
 throughout this paper.

\section{Observations and Data Reduction}

  HIJASS is the northern 
 counterpart to the HI Parkes All Sky Survey (HIPASS) \citep{sta96}
 which,  when complete, will have surveyed the 
 whole sky up to $\delta$=25\degr. HIJASS will survey 
 the northern sky above this declination to similar 
 sensitivity. HIJASS is being undertaken
 using the Multibeam 4-beam cryogenic receiver mounted on the 76-m Lovell 
 telescope (beam FWHM $\simeq$12.0\arcmin) at Jodrell Bank. A 64~MHz  bandpass with 1024 
 channels  is used, although local interference restricts the useful 
  velocity range to about --1000~km\,s$^{-1}$ to +10,000~km\,s$^{-1}$. 
  The survey is conducted by scanning the receiver in
  declination strips of $\simeq$8\degr. Each declination strip is separated by 10\arcmin\, 
  but each area of sky is scanned 8 times, resulting in a final scan separation of
   1.25\arcmin.   
 Bandpass correction and calibration  
 are applied using the same software as HIPASS \citep{bar01}. 
   The spectra are gridded into three-dimensional 
   8\degr$\times$8\degr\,  
 data cubes ($\alpha$,$\delta$,$V_{\odot}$) which have an rms noise level of 
 $\simeq$16~mJy~beam$^{-1}$.  The
  FWHM velocity resolution is 18.0~km\,s$^{-1}$ and the spatial 
  pixel size  is 4\arcmin$\times$4\arcmin.

 To survey the M81 group we searched the HIJASS data  between  $V_{\odot}$=$\pm$500~km\,s$^{-1}$
  in the regions 8$^{h}$$<$$\alpha$$<$11$^{h}$, +70$^\circ$$<$$\delta$$<$+78$^\circ$ 
  and  
  9$^{h}$03$^{m}$$<$$\alpha$$<$10$^{h}$36$^{m}$, +62$^\circ$$<$$\delta$$<$+70$^\circ$. 
 This area ($\simeq$180~sq~deg) 
 includes the whole of the HI complex around M81 and most of the outer area of the 
 group. 
 The HIJASS data was hanning smoothed 
 by 3 channels. This reduces the velocity resolution to 26~km\,s$^{-1}$
 but improves the rms noise to $\simeq$11~mJy~beam$^{-1}$. 
   The relatively high spatial resolution and full spatial sampling of HIJASS  
  make it more effective at revealing dwarf galaxies within 
  the HI complex around M81 than previous pointed  surveys. 
  HIJASS is more sensitive to low column density gas than aperture synthesis 
  surveys.




\section{Results}

Fig.~1 is a set of R.A.--Decl. plots of the HI
   emission from the HI complex around M81 at eight selected velocities.
   Fig.~2 is a set 
 of Velocity--Decl. plots at five selected R.A.s.

\begin{figure}
\epsscale{0.8}
\plotone{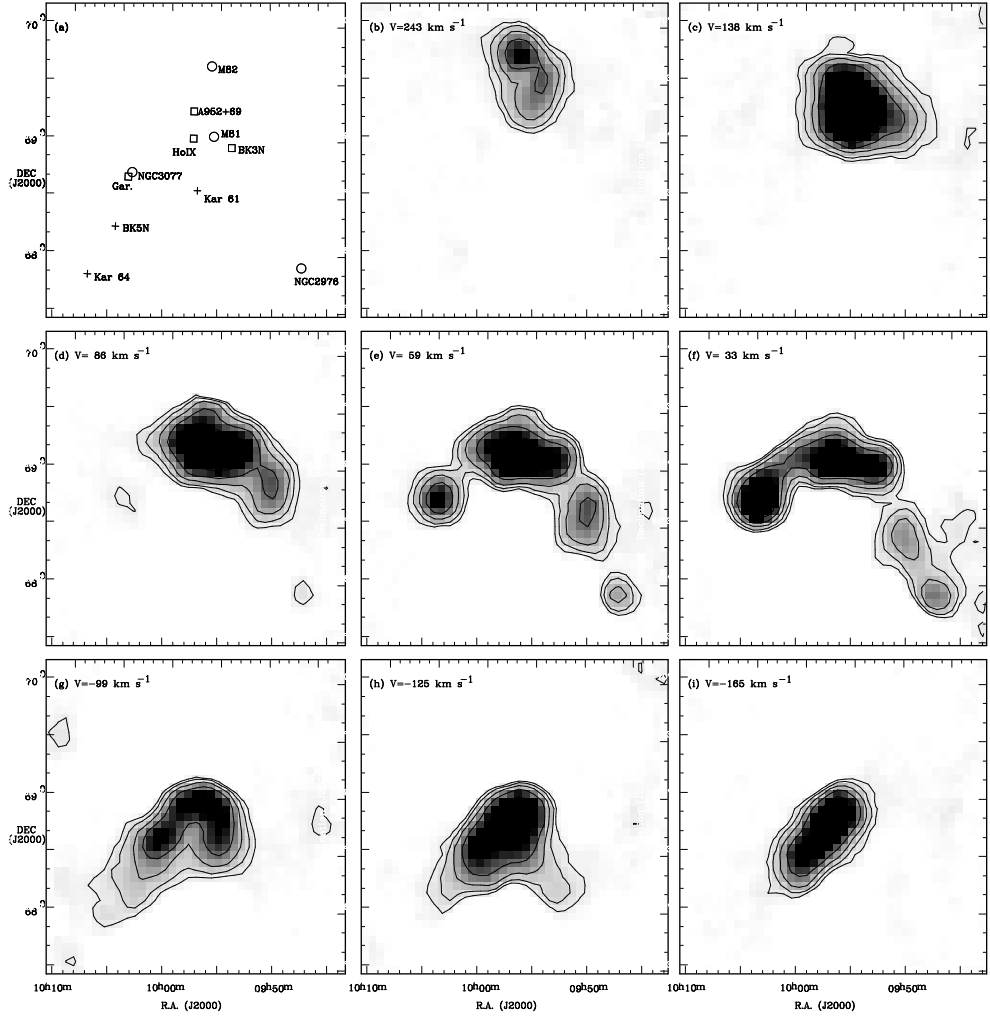}
\caption{R.A.-Decl. plots at selected velocity 
 channels from hanning smoothed HIJASS data of the area around the M81 HI complex. 
 The contours are set at 0.05, 0.1, 0.2, 0.4, 0.8, 1.6, 3.2 mJy~beam$^{-1}$. 
 (a) positions of known galaxies in area (circles denote giant galaxies, 
 squares show dIrrs, crosses show dEs), (b) $V_{\odot}$=243~km\,s$^{-1}$, (c) $V_{\odot}$=138~km\,s$^{-1}$, (d) $V_{\odot}$=86~km\,s$^{-1}$, (e) $V_{\odot}$=59~km\,s$^{-1}$,
  (f) $V_{\odot}$=33~km\,s$^{-1}$, (g) $V_{\odot}$=--99~km\,s$^{-1}$, 
 (h) $V_{\odot}$=--125~km\,s$^{-1}$, (i) $V_{\odot}$=--165~km\,s$^{-1}$.}
\end{figure}

\begin{figure}
\epsscale{0.4}
\plotone{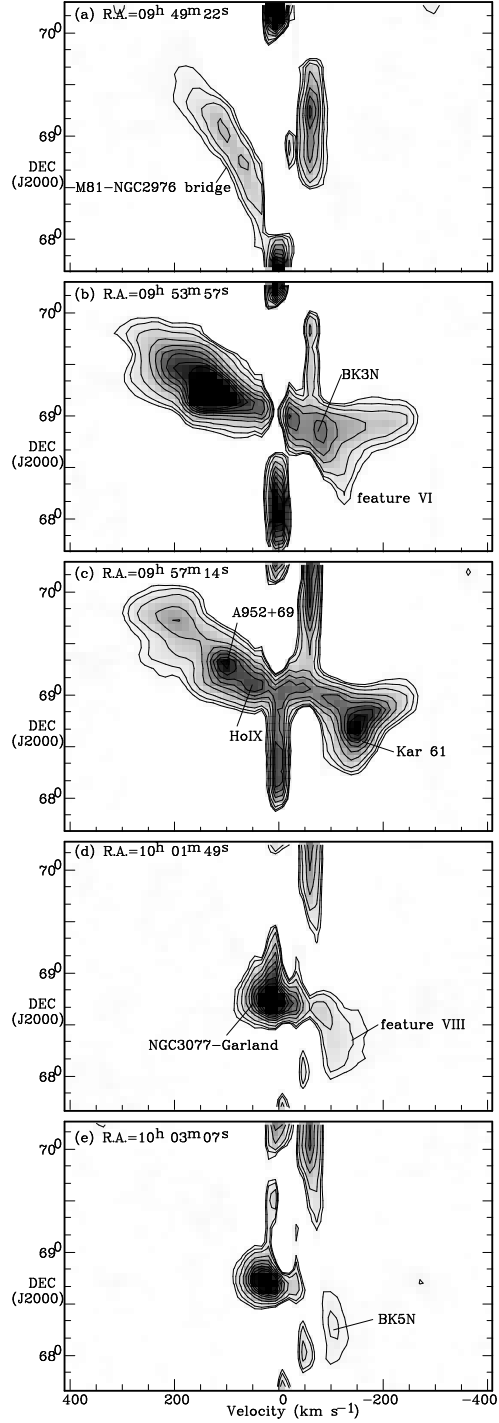}
\caption{ Velocity-Decl. plots 
 at selected R.A.s from hanning smoothed HIJASS data of the area around the M81 HI complex. 
  The contours are set at 0.05, 0.1, 0.2, 0.4, 0.6, 0.8, 1.0, 1.2, 1.4, 1.6 mJy~beam$^{-1}$.
 (a) R.A.=09$^h$49$^m$22$^s$, (b) R.A.=09$^h$53$^m$57$^s$,
  (c) R.A.=09$^h$57$^m$14$^s$, (d) R.A.=10$^h$01$^m$49$^s$, 
 (e) R.A.=10$^h$03$^m$07$^s$. }
\end{figure}

The tidal bridge of gas between M81 and M82 
 \citep[see e.g.][]{app81,yua94} can be clearly seen in the HIJASS data, 
  for example in Fig.~1b at $\alpha$$\simeq$9$^h$54$^m$,
 $\delta$$\simeq$69\degr30\arcmin\, at 
  $V_{\odot}$=243~km\,s$^{-1}$ and  
        in Fig.~2c stretching from 
   $\delta$$\simeq$69\degr45\arcmin, $V_{\odot}$$\simeq$250~km\,s$^{-1}$ 
 to  $\delta$$\simeq$69\degr00\arcmin, $V_{\odot}$$\simeq$0~km\,s$^{-1}$. 
 Within Fig.~2c can also be clearly seen the dIrr galaxies Holmberg~IX 
 (at $\delta$$\simeq$69\degr03\arcmin, $V_{\odot}$$\simeq$50~km\,s$^{-1}$) 
 and A952+69 (at  $\delta$$\simeq$69\degr15\arcmin, 
  $V_{\odot}$$\simeq$100~km\,s$^{-1}$). A952+69 has not been unambiguously discerned in 
  single-dish surveys \citep[see][]{vd98} although both 
 objects are seen in the VLA data of \citet{yua94} who labelled 
 them `concentration~I' (Holmberg~IX) and `concentration~II' (A952+69). 
 The HIJASS data (Fig.~2c) show them to form part of the tidal bridge between 
 M81 and M82. We suggest that both  galaxies may have  recently condensed 
  from the tidal debris between M81 and M82. 
  Morphologically such `tidal dwarf galaxies' are very similar to classical dIrrs 
  \citep[][]{wei00}.

 A bridge of gas apparently connecting  NGC~3077 and 
  M81  can be clearly  seen at 
 $\alpha$$\simeq$10$^h$00$^m$, $\delta$$\simeq$68\degr50\arcmin, 
  $V_{\odot}$=33~km\,s$^{-1}$ (Fig.~1f).  
 \citet{yua94} dubbed this the `north tidal bridge' and 
  considered it to result
 from a tidal interaction between M82 and NGC3077. 
 The gas around NGC~3077 is actually centred a few arcmins west of 
  the galaxy and $\simeq$55km\,s$^{-1}$ higher in velocity.  
 The dIrr galaxy Garland lies approximately at the centre of 
 the gas. 
   \citet{vd98} suggested  that Garland may be at an intermediate 
 stage in the conversion of a tidal tail into a dwarf galaxy.

 Fig.~1 reveals a bridge of gas stretching from M81 at 
 $\alpha$$\simeq$09$^h$52$^m$ $\delta$$\simeq$69\degr00\arcmin, 
  $V_{\odot}$=138~km\,s$^{-1}$, (Fig.~1c) to 
 NGC2976 at  
 $\alpha$$\simeq$9$^h$48$^m$, $\delta$$\simeq$67\degr50\arcmin, 
 $V_{\odot}$=33~km\,s$^{-1}$ (Fig~1f).
  \citet{app81} considered this to 
  consist of three linked features. However, 
 the HIJASS data show that it forms a single tidal bridge between the 
 galaxies (see Fig.~2a at R.A.=09$^h$49$^m$22$^s$). 
 The bridge has an HI mass of $\simeq$2.1$\times$10$^{8}$~M$_{\odot}$ 
  and  projected 
 spatial extent of $\simeq$80~kpc. 

 The spur of HI emission labelled `feature VIII' by Appleton et al. can be 
 traced over a much larger spatial and velocity range in the HIJASS data. 
  Its full spatial extent can be seen in 
 Fig.~1g ($V_{\odot}$$\simeq$--99~km\,s$^{-1}$) where it extends 
   from close to NGC3077 out to   
  $\alpha$$\simeq$10$^h$05$^m$, $\delta$$\simeq$67\degr45\arcmin, 
 a projected spatial size of $\simeq$75~kpc. 
    The spur has a HI mass of 
  $\simeq$3.1$\times$10$^8$~M$_{\odot}$.  
 The spur is strikingly seen in Fig.~2d (at R.A.=10$^h$01$^m$49$^s$), 
 stretching from $\delta$$\simeq$68\degr35\arcmin, $V_{\odot}$$\simeq$0~km\,s$^{-1}$ 
 to $\delta$$\simeq$67\degr50\arcmin, $V_{\odot}$--100~km\,s$^{-1}$.  
  The apparent association of the high declination end of the 
 spur with NGC3077-Garland is clear from this. 
 The lower declination part of the spur continues to be seen in Fig.~2e (at 
 R.A.=10$^h$03$^m$07$^s$). The peak in the emission at 
 $\delta$$\simeq$68\degr15\arcmin, 
   $V_{\odot}$$\simeq$--100~km\,s$^{-1}$ occurs at the 
 optical position of the dE galaxy BK5N. There is the suggestion of a second peak 
 in the HI emission at $\delta$$\simeq$67\degr54\arcmin, 
 $V_{\odot}$$\simeq$--100~km\,s$^{-1}$. 
 This peak can be seen out to R.A.$\simeq$10$^h$05$^m$, very close to the 
 optical position of the nucleated dE galaxy Kar~64. 
  We suggest that the spur may 
      result from tidal interaction
   between NGC3077 and either or both of BK5N and Kar~64.

 Kar~61, the closest dE to M81 itself, can be  strikingly seen  
  in Fig.~2c (at R.A.=09$^h$57$^m$14$^s$)
   at $\delta$$\simeq$68\degr40\arcmin, $V_{\odot}$$\simeq$--140~km\,s$^{-1}$. 
 Whilst previous HI maps had shown emission at the position of this 
 galaxy, it had not been clear whether this emission resulted from Kar~61 or 
 from M81 \citep[see][]{vd98}. 
  \citet{joh97} found a bright HII region knot situated NE of the 
 galaxy centre from which they measured  a radial
    velocity 
 of --135$\pm$30~km\,s$^{-1}$. 
 The correspondence of this velocity 
 with that of the HI peak  confirms that the HI is associated with 
 this object. We estimate that at least $\sim$10$^{8}$~M$_{\odot}$ 
  of HI is associated with 
   Kar~61.  On the basis of HST imaging, 
 \citet{kar00} suggested that Kar~61 may be  a dE/dIrr transition type.  
 The presence of a large amount of HI supports this idea.  

 The spur of HI emission labelled `feature~VI' by Appleton et al. is traced
  over a larger spatial and velocity extent in the HIJASS data. It can be clearly 
 seen in Fig.~1h at  $\alpha$$\simeq$9$^h$51$^m$, 
 $\delta$$\simeq$68\degr0\arcmin\, at $V_{\odot}$$\simeq$--125~km\,s$^{-1}$ (Fig.~1h)
 and in  
  Fig.~2b (at R.A.=09$^h$53$^m$57$^s$) stretching out to $\delta$$\simeq$68\degr15\arcmin, V$\simeq$--125~km\,s$^{-1}$. 
   This spur has a projected spatial extent of $\simeq$50~kpc and an HI mass of 
  $\simeq$6.5$\times$10$^{7}$~M$_{\odot}$.  
   The dIrr galaxy BK3N 
 appears to lie at end of this spur closest to M81. This 
 object is best seen in Fig.~2b at  
 $\delta$$\simeq$68\degr50\arcmin, $V_{\odot}$$\simeq$--60~km\,s$^{-1}$. 
 This is the first unambiguous HI detection of this object, since previous
 surveys were unable to disentangle its emission from that of M81 
 \citep[see][]{vd98}. We estimate $\simeq$1.8$\times$10$^{7}$~M$_{\odot}$ of HI is  
associated with the galaxy. 
 BK3N's position at the 
top of the HI-spur suggests that it may be a `tidal dwarf galaxy', condensing from 
 the tidal debris of the spur. Alternatively, BK3N may be a pre-existing
  dIrr galaxy undergoing an interaction with M81. 

There are within our survey area, but outside the M81 HI 
complex, a further 5 optically identified dIrr group members 
 (Holmberg II, Kar~52, UGC4483, Holmberg I, Kar~73). All 
 of these had been previously detected in HI. 
The HIJASS 
 survey confirms these detections.
 The BCD galaxy UGC5423,  
  which had also been previously detected in HI, is also 
 detected by HIJASS. The other BCD (DDO82) in our survey 
 area is not detected by HIJASS. We do not detect
 any of the dE galaxies which lie within our survey area but outside the 
  M81 HI complex (i.e. Kar~74, DDO71,  BK6N, 
 F8D1, KK77, FM1, KKH57).

\begin{figure}
\epsscale{0.8}
\plotone{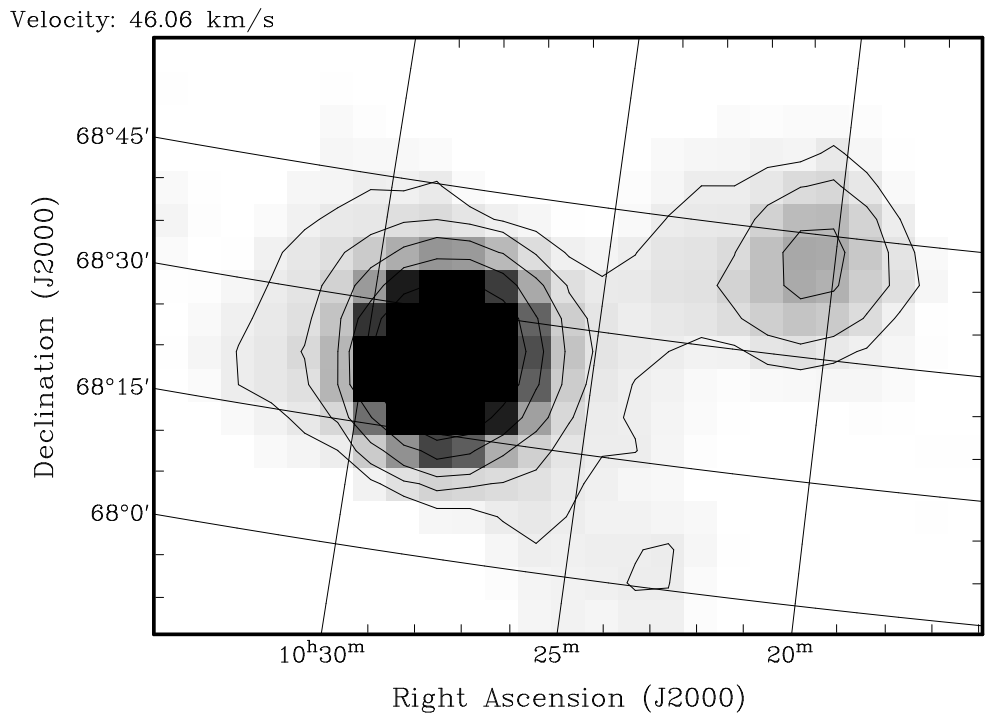}
\caption{R.A.-Decl plot at $V_{\odot}$=46~km\,s$^{-1}$ showing the 
 HI associated with IC2574 (left) and the newly discovered object
 HIJASS~J1021+6840.  The contours are set at 0.05, 0.1, 0.2, 0.4, 0.8, 1.6, 3.2 mJy~beam$^{-1}$.}
\end{figure}

An extensive search of the HIJASS data 
 for previously unidentified M81 group members revealed only 
 one strong candidate, HIJASS~J1021+6842. 
 Fig.~3 is a  R.A.-Decl. plot at $V_{\odot}$=46~km\,s$^{-1}$    
  showing this object. It lies $\simeq$105\arcmin\, from  
  IC2574 (projected separation $\simeq$112~kpc) and there 
 is marginal evidence for a connection to this galaxy. 
 Its FWHM Velocity is $\simeq$50~km\,s$^{-1}$ and its HI 
 mass is  $\simeq$3$\times$10$^{7}$~M$_{\odot}$ 
 if it is a member of the M81 group. No optical counterpart can be 
 seen on the 2nd generation red Digital Sky Survey.  
   It may be a very low surface brightness
   dIrr companion to IC2574.  
  We may be  seeing the last remnants of a tidal encounter 
  between IC2574 and one of the galaxies around M81.

\section{Discussion and Conclusions}

 The M81 group shows a clear morphology-density relation:  
 dEs are found in and around the dense core close to M81, while 
   dIrrs are spread over a much larger area. 
 Dwarf galaxies in the Local Group 
 exhibit a similar relation  \citep{gre00}. 
  N-body simulations \citep{may01} 
 suggest that a  close tidal interaction of a dIrr with a giant galaxy can 
 induce severe mass loss and non-axisymmetric instabilities in the disk of the 
 dIrr, 
  turning it into an object which matches the observed properties of a dE. 
  The implication is that dIrrs close to the Milky Way or M31 have been 
  transformed in this way into the dEs now observed. 
 Those dIrrs further away from the giants  have not yet undergone  
  a close tidal encounter and remain recognisable as dIrrs.
 The HIJASS data provide some evidence that  
  a similar  scenario may   explain the morphological 
 segregation of the dwarf galaxies in the M81 group. 
 
    The dEs
   BK5N and Kar~64 appear to be associated with  
   a tidal spur of HI mass $\simeq$3.1$\times$10$^{8}$~M$_{\odot}$, 
 stretching from 
   NGC3077 out a projected distance of $\simeq$75~kpc. 
   We suggest that BK5N and Kar~64 may be former dIrrs 
  which have been stripped of their gas by tidal encounters with NGC3077.

 Kar~61, the only dE galaxy closer to M81 than BK5N,  
   has $\sim$10$^{8}$~M$_{\odot}$ of HI associated with it, 
 shows signs of recent star formation and has been described as 
  a dE/dIrr transition object. We suggest that Kar~61  
  is involved in a tidal interaction with M81 but is at a much earlier 
 stage of tidal stripping than BK5N or Kar~64. 
  We did not  detect any of the  
  dEs which lie further from  M81  than Kar~61, BK5N and Kar~64. 
  These dEs may already have had all of their HI stripped 
  by previous tidal encounters. 

   How have the four known dIrrs 
 within the M81 HI complex  survived if other dIrrs 
 have been tidally stripped and transmuted into dEs ?  
  At least three of these 
 (Holmberg IX, A952+69, Garland)   
  may actually be tidal dwarf galaxies. The fourth, 
 BK3N, may also be a tidal dwarf galaxy or it may be 
 a  dIrr currently being tidally stripped. 
 The newly discovered object HIJASS~J1021+6842 may be a further example 
 of a tidal dwarf galaxy in formation, in this 
  case one in which star formation has not yet begun.



\acknowledgements

PJB and RFM acknowledge the financial support of the UK PPARC.
 This research has made use of the NASA/IPAC Extragalactic Database (NED) 
 which is operated by the Jet Propulsion Laboratory, Caltech, under agreement 
 with the National Aeronautics and Space Administration.




\end{document}